# TaAs Weyl semimetal based one-dimensional photonic structures

Ilka Kriegel[1], Michele Guizzardi[2], Francesco Scotognella[2,3,*]

[1]*Department of Nanochemistry, Istituto Italiano di Tecnologia (IIT), via Morego, 30, 16163 Genova, Italy*
[2]*Dipartimento di Fisica, Politecnico di Milano, Piazza Leonardo da Vinci 32, 20133 Milano, Italy*
[3]*Center for Nano Science and Technology@PoliMi, Istituto Italiano di Tecnologia, Via Giovanni Pascoli, 70/3, 20133, Milan, Italy*
*email address: francesco.scotognella@polimi.it

**Abstract**
Weyl semimetals can be described as the three-dimensional analogue of graphene, showing linear dispersion around nodes (Weyl points) [1]. Tantalum arsenide (TaAs) is among the most studied Weyl semimetals. It has been demonstrated that TaAs has a very high value of the real part of the complex refractive index in the infrared region [2]. In this work we show one-dimensional photonic crystals alternating TaAs with $SiO_2$ or $TiO_2$ and a microcavity where a layer of TaAs is embedded between two $SiO_2$-$TiO_2$ multilayer.

**Introduction**
Tantalum arsenide (TaAs) is a deeply studied Weyl semimetal. Angle resolved photoemission spectroscopy measurements together with *ab initio* calculations demonstrated that fermions in TaAs can be described as massless chiral particles with spin ½ [3–7]. Recently, Wu et al. [8] have measured a very intense nonlinear optical response of TaAs together with other monopnictide Weyl semimetals, while Weber et al. [9] have studied the ultrafast dynamics of TaAs. Buckeridge et al. [2] have calculated several properties of TaAs, including the dielectric function of the material. Owing to the tetragonal symmetry, an anisotropy of the optical properties arises, with two different dielectric functions when the oscillating electric field is parallel to the *a* and *b* axis or parallel to the *c* axis. In the *c* direction, at photon energies below 1 eV, the imaginary part of the refractive index decreases significantly, while the real part of refractive index has values above 6. Thus, TaAs can be exploited as a high refractive index material in the infrared region.

For this reason we want to demonstrate in this work the implementation of tantalum arsenide as layer in a one-dimensional (1D) multilayer photonic crystal. 1D photonic crystals are the simplest case of photonic crystals that are materials in which the alternation of high and low refractive index has a length scale of the light wavelength. In these materials light is not allowed to propagate for certain photon energies, in correspondence of the so-called photonic band gap [10–13]. 1D multilayer photonic crystals can be fabricated with many fabrication techniques as sputtering, spin coating, pulsed laser deposition, chemical etching and molecular beam epitaxy [14–18]. Herein we show one-dimensional photonic crystals alternating TaAs with $SiO_2$ or $TiO_2$ and a microcavity where a layer of TaAs is embedded between two $SiO_2$-$TiO_2$ multilayers. We show that with very thin layers of TaAs, 16 nm in the photonic crystals and 8 nm in the microcavity, we can achieve very efficient photonic band gaps and cavity defects, due to the high real part of the refractive index of TaAs in the infrared region.

**Methods**
To calculate the optical properties of the different photonic structures studied herein we have employed the transfer matrix method [19,20]. In our study the light impinges at normal incidence a system composed of a glass substrate, the photonic structure and air. We have studied the optical response of a TaAs/SiO$_2$ photonic crystal, of a TaAs/TiO$_2$ photonic crystal and of a (TiO$_2$-SiO$_2$)$_7$-[SiO$_2$-TaAs-SiO$_2$]-(SiO$_2$-TiO$_2$)$_7$ microcavity in the range 0.2 – 1.1 eV with a step of 2.5 meV.

**Results and Discussion**
We study in this work the optical properties of 1D photonic crystals made with TaAs alternated with SiO$_2$ or TiO$_2$ and of a microcavity in which TaAs is embedded between two SiO$_2$-TiO$_2$ Bragg mirrors (7 bilayers), as sketched in Figure 1.

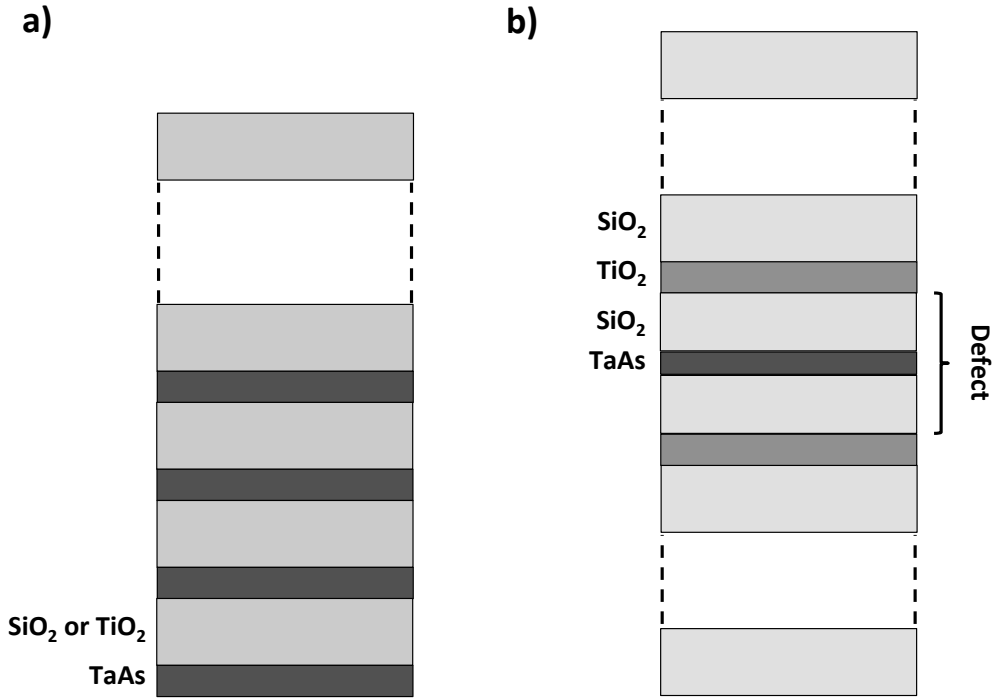

**Figure 1.** Scheme of (a) a TaAs/SiO$_2$ (or TaAs/TiO$_2$) photonic crystal and of (b) a (TiO$_2$-SiO$_2$)$_7$-[SiO$_2$-TaAs-SiO$_2$]-(SiO$_2$-TiO$_2$)$_7$ microcavity.

In Figure 2 we show the light transmission spectrum of a photonic crystal made of 7 bilayers of TaAs and SiO$_2$. The thickness of each TaAs layer is 16 nm, while the thickness of each SiO$_2$ layer is 1200 nm. $n$ and $k$ (real part and imaginary part of the refractive index) of TaAs, in the $c$ direction of the crystal, have been taken from [2]. The Sellmeier equation for the dispersion of the refractive index of silica is [21]:

$$n^2_{SiO_2}(\lambda) - 1 = \frac{0.6961663\lambda^2}{\lambda^2 - 0.0684043^2} + \frac{0.4079426\lambda^2}{\lambda^2 - 0.1162414^2} + \frac{0.8974794\lambda^2}{\lambda^2 - 9.896161^2} \qquad (1).$$

The Sellmeier equation for the dispersion of the refractive index of titania, reliable for thin films in the range 0.2 – 1.1 eV, fits the data by Siefke et al. [22,23] and is

$$n^2_{TiO_2}(\lambda) - 1 = \frac{4.181\lambda^2}{\lambda^2 - 0.2133^2} + \frac{5.068\lambda^2}{\lambda^2 - 14.33^2} \qquad (2).$$

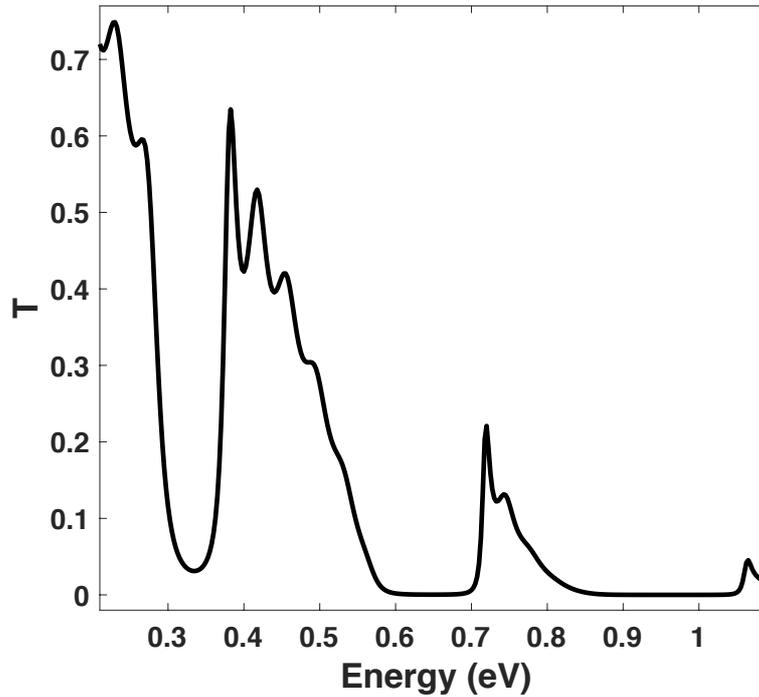

**Figure 2.** Light transmission spectrum of a 1D photonic crystal made by 7 bilayers of TaAs and $SiO_2$.

For thickness of TaAs above 16 nm the light absorption of the material becomes very strong because of the high imaginary part of the refractive index.

If we use $TiO_2$ instead of $SiO_2$ to be alternated with TaAs, we observe a photonic band gap in the same spectral position (i.e. 0.35 eV) with a thickness of the $TiO_2$ layers of 765 nm, resulting in a thinner photonic crystal. In Figure 3 the light transmission spectrum of a photonic crystal made of TaAs and $TiO_2$ is depicted.

In the spectra in Figure 2 and 3 the transmission valley around 0.65 eV is the second order of the photonic band gap of the photonic crystals.

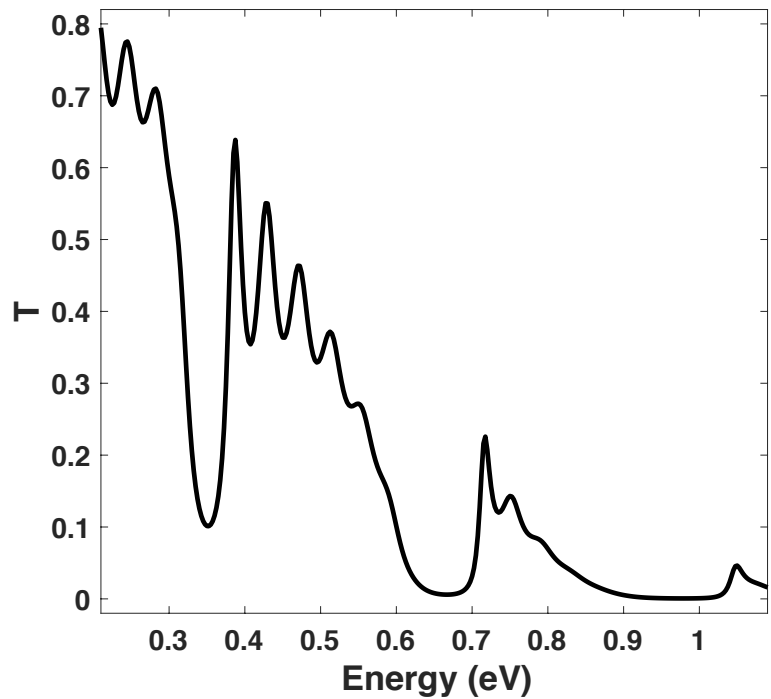

**Figure 3.** Light transmission spectrum of a 1D photonic crystal made by 7 bilayers of TaAs and $TiO_2$.

The photonic band gap is less intense with TaAs-$TiO_2$ unit cell (minimum at about 0.1) with respect to the one with TaAs-$SiO_2$ unit cell (minimum at about 0.03) because of the smaller refractive index contrast in the TaAs-$TiO_2$ case.

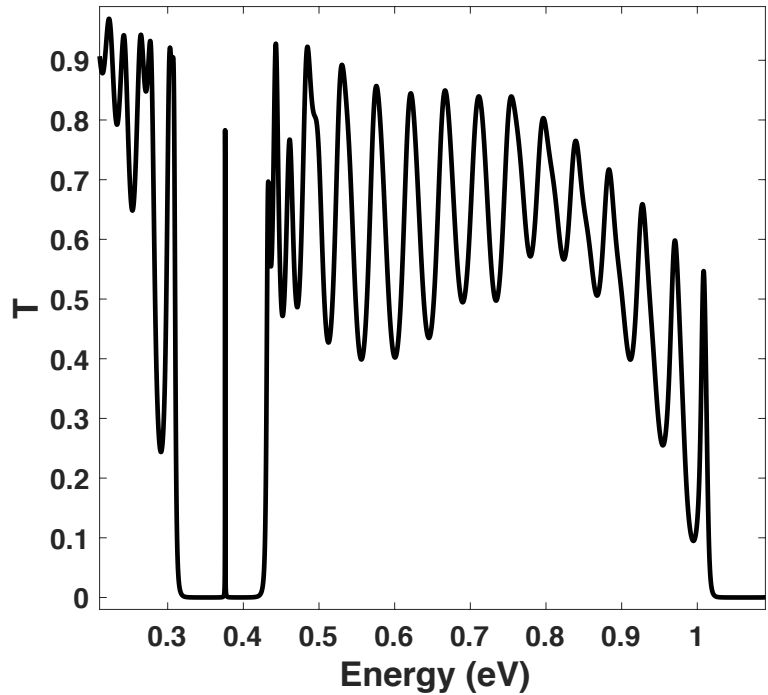

**Figure 4.** Light transmission spectrum of a $(TiO_2$-$SiO_2)_7$-$[SiO_2$-TaAs-$SiO_2]$-$(SiO_2$-$TiO_2)_7$ microcavity.

In Figure 4 we show the light transmission spectrum of a $(TiO_2$-$SiO_2)_7$-$[SiO_2$-TaAs-$SiO_2]$-$(SiO_2$-$TiO_2)_7$ microcavity. The $TiO_2$ layers of the $[TiO_2$-$SiO_2]_7$ Bragg mirrors are 377.78 nm thick, while the $SiO_2$ layers are 590.28 nm thick. The two $SiO_2$ layers that sandwich the TaAs layer in the cavity defect are 560 nm thick, while the thickness of the TaAs layer is 8 nm. The microcavity shows in the region 0.31 – 0.42 eV a photonic band gap with an intense defect peak at 0.376 eV.

**Conclusion**
In this work we have exploited the high refractive index of tantalum arsenide in the infrared region (in the region 0.2 – 1.1 eV) employing the material as layer in the engineering of a 1D multilayer photonic crystal and of a 1D microcavity. We show one-dimensional photonic crystals alternating TaAs with $SiO_2$ or $TiO_2$ and a microcavity where a layer of TaAs is embedded between two $SiO_2$-$TiO_2$ multilayers. It is worth noting that with thin TaAs layers, 8 to 16 nm thick, it is possible to build efficient photonic crystals and microcavities in the infrared region with good efficiency. This can be promising for the fabrication of thin filters for this spectral region.

**Acknowledgement**
This project has received funding from the European Union's Horizon 2020 research and innovation programme (MOPTOPus) under the Marie Skłodowska-Curie grant agreement No. [705444], as well as (SONAR) grant agreement no. [734690].